# Universal quantum oscillations in the underdoped cuprate superconductors

Neven Barišić[1,2,3], Sven Badoux[4], Mun K. Chan[2], Chelsey Dorow[2], Wojciech Tabis[2], Baptiste Vignolle[4], Guichuan Yu[2], Jérôme Béard[4], Xudong Zhao[2,5], Cyril Proust[4], Martin Greven[2]

[1] Service de Physique de l'Etat Condensé, CEA-DSM-IRAMIS, F 91198 Gif-sur-Yvette, France

[2] School of Physics and Astronomy, University of Minnesota, Minneapolis, Minnesota 55455, USA

[3] Institute of Physics, HR–10000 Zagreb, Croatia

[4] Laboratoire National des Champs Magnétiques Intenses, UPR 3228, (CNRS-INSA-UJF-UPS), Toulouse 31400, France.

[5] State Key Lab of Inorganic Synthesis and Preparative Chemistry, College of Chemistry, Jilin University, Changchun 130012, China.

**The metallic state of the underdoped high-$T_c$ cuprates has remained an enigma: How may seemingly disconnected Fermi surface segments, observed in zero magnetic field as a result of the opening of a partial gap (the pseudogap), possess conventional quasiparticle properties[1-3]? How do the small Fermi-surface pockets evidenced by the observation of quantum oscillations (QO) emerge as superconductivity is suppressed in high magnetic fields[4-11]? Such QO, discovered in underdoped YBa$_2$Cu$_3$O$_{6.5}$ (Y123)[12] and YBa$_2$Cu$_4$O$_8$ (Y124)[13], signify the existence of a conventional Fermi surface (FS)[3,14,15]. However, due to the complexity of the crystal structures of Y123 and Y124 (CuO$_2$ double-layers, CuO chains, low structural symmetry), it has remained unclear if the QO are specific to this particular family of cuprates[5,16-18]. Numerous theoretical proposals have been put forward to explain the route toward QO, including materials-specific scenarios involving CuO chains and scenarios involving the quintessential CuO$_2$ planes[6-11,14-18]. Here we report the observation of QO in underdoped HgBa$_2$CuO$_{4+\delta}$ (Hg1201), a model cuprate superconductor with individual CuO$_2$ layers, high tetragonal symmetry, and no CuO chains (Fig. 1). This observation proves that QO are a universal property of the underdoped CuO$_2$ planes, and it opens the door to quantitative future studies of the metallic state and of the Fermi-surface reconstruction phenomenon in this structurally simplest cuprate.**

There exists a broad consensus that the cuprates approach conventional metallic behavior at very high hole-dopant concentrations ($p$)[19-21]. For example, Tl$_2$Ba$_2$CuO$_{6+\delta}$ at $p \approx 0.30$ is characterized by a large hole-like FS that corresponds to approximately 65% of the first Brillouin zone (containing $1+p$ carriers per CuO$_2$ plane), in good agreement with band structure calculations[15,20,21]. This is contrasted by the yet unexplained high-temperature metallic state at intermediate hole concentrations, and by the strong insulating behavior in the absence of doped carriers. The observation for Y123[12] and Y124[13] of QO not far from the undoped insulating state (for $p \approx 0.1$) is surprising, as QO are a prominent characteristic of simple conventional metals, and it has triggered much theoretical activity[6-11,14-18]. Moreover, the observed oscillation period implies a very small pocket (covering only ~2% of the zone), in stark contrast to the situation at high hole-dopant concentrations. While this result can been taken as evidence for a dramatic change of the FS associated with the quintessential $CuO_2$ planes, it may also be attributed to the existence of a non-universal FS piece related to the CuO chains in Y123 and Y124[5,16-18].

In the latter scenario, hybridization between CuO chain and BaO plane states gives rise to a small hole pocket[16-18]. This scenario seems to be supported by the observation of a small residual electronic term of the specific heat, and by the persistence of the $\sqrt{H}$ signature of a $d$-wave superconducting gap to high fields, even upon the appearance of QO, indicating the coexistence of two very different types of Fermi surfaces[5]. Consequently, the superconducting properties may be attributed to the large hole FS due to the $CuO_2$ planes, whereas the QO may be associated with pockets present already at low fields. Additional effects due to an electronic instability of the CuO chains may occur, akin to well-known examples in organic metals where quasi-one-dimensional bands can support spin-density wave (SDW) order induced by a magnetic field, resulting in QO related to the reconstructed closed FS[22].

The other class of scenarios invokes symmetry breaking within the $CuO_2$ planes that leads to FS reconstruction at low temperature. Indeed, high-field nuclear magnetic resonance measurements revealed static charge-density modulations within the $CuO_2$ planes of underdoped Y123[23]. Subsequent X-ray studies of Y123 found charge-density-wave (CDW) correlations[24,25], with comparable correlation length along and perpendicular to the CuO chains[25], consistent with the notion that the QO are a property of the $CuO_2$ planes. However, the orthorhombic structure and the chain layers impose a special direction in Y123 and Y124, which might promote and stabilize the charge modulation. Consequently, it has remained a pivotal open question whether the FS reconstruction has anything to do with aspects of the unidirectional structures, or if it is a universal property of the cuprates.

Here we settle this issue through a transverse magnetorestivity study of Hg1201 at $p \approx$ 0.09 in pulsed magnetic fields of up to 80 T. This doping level is close to that for which QO were initially observed in Y123. Only samples with sufficiently low scattering rate can show the passage of the Landau levels across the Fermi level. A major obstacle to overcome was the synthesis and electrical contacting of ultra-pure single crystals with negligible residual resistivity[1]. This was achieved according to previously reported protocols[26,27] - see also Supplementary Information (SI).

Figure 2 shows isothermal magnetoresistivity curves for four different temperatures above the irreversibility field ($H_{irr}$ = 20-30 T). The applied current was parallel to the $CuO_2$ planes, while the magnetic field was perpendicular to the planes. The oscillatory behavior, which is our main result, is already evident from the raw data at fields above 60 T. In order to emphasize the QO without "background" subtraction, we also plot the derivative of the data (inset of Fig. 2). We note that the oscillations at all four temperatures are in phase with each other. Quantum oscillations with the same frequency were observed in another sample with similar $T_c$, but with lesser precision (see SI).

In Fig. 3a, a smooth non-oscillatory contribution is removed and the result is plotted versus $1/B$. The Fourier transforms in the limited field range [62 T; 81 T] exhibit a single peak at $F = 840 \pm 30$ T (Fig. 3b). According to the Onsager relation, $F = A_k \Phi_0 / 2\pi^2$, where $\Phi_0$ is the magnetic flux quantum and $A_k$ the cross-sectional area of the FS perpendicular to the applied field, which corresponds to about 3 % of the Brillouin zone. Assuming that the FS is strictly two-dimensional, the Luttinger theorem yields $n_{2D} = 2A_k/(2\pi)^2 = F/\Phi_0 = 0.061 \pm 0.002$ carriers per pocket. In a single-band model, the Hall coefficient $R_H = 1/n_{3D}e$ ($n_{3D} = n_{2D}/c$, where $c$ is the lattice parameter perpendicular to the $CuO_2$ planes) can be evaluated to be $|R_H| = 14.7 \pm 0.6$

mm$^3$/C, in very good agreement with the value |$R_H$| = 15 ± 5 mm$^3$/C obtained at low temperatures and in high fields for a Hg1201 sample with a similar doping level ($T_c$ = 65 K, $p$ ≈ 0.075)[28]. By following the temperature dependence of the oscillation amplitude (inset of Fig. 3b), the quasiparticle effective mass $m^*$ = 2.45 ± 0.15 $m_e$ is extracted (where $m_e$ is the free-electron mass). By performing a Lifshitz-Kosevich fit (solid lines in Fig. 3a), the Dingle temperature can be evaluated to be $T_D$ = (18 ± 4) K, which corresponds to a mean free path of $\ell$ ≈ 5 nm.

The QO in Y123 and Hg1201 appear approximately in the same doping, temperature and magnetic field range of the phase diagram, with pockets of very similar effective mass and comparable cyclotron frequency, as summarized in Table 1. Moreover, the temperature dependence and absolute value of the Hall and Seebeck coefficients are nearly the same for $p$ ≈ 0.09. Remarkably, upon cooling, both coefficients sharply decrease below 50 K, and finally change sign from positive to negative at low temperature[28]. This supports the scenario of a FS reconstruction that leads to a small, closed and coherent electron pocket, and it points to the same underlying mechanism related to the CuO$_2$ plane, since this is the only common structural unit of the two compounds (Fig. 1). In Y123, the FS reconstruction is consistent with bi-axial CDW order that competes with superconductivity[14,23-25]. Our data for Hg1201 do not allow us to extract the number of pockets, although it appears likely that there exists one pocket per CuO$_2$ plane.

A recent analysis that spanned a broad doping and temperature range for a number of cuprates, including Y123 and Hg1201, determined the universal zero-field sheet resistance associated with the nodal states[1]. Below the pseudogap temperature $T^*$, where neutron diffraction measurements show the onset of an unusual magnetic state in both Y123 and Hg1201[29], it was demonstrated that there exists another characteristic temperature ($T^{**}$) below which the planar transport becomes Fermi-liquid like[1-3]. For Y123, the temperature below which the CDW correlations are found approximately coincides with $T^{**}$. Although CDW correlations remain to be reported for Hg1201, the similarity of the transport properties of the two compounds is remarkable. A central question is the relationship between the FS reconstruction suggested by quantum oscillation measurements, the CDW correlations observed already in the zero-field high-temperature Fermi-liquid-like state, and the opening of the pseudogap at $T^{*}$[8,9]. Our results rule out theories that relate the reconstruction of the FS to a unidirectional character of the crystallographic structure, and they imply that the applicability of theoretical models based on the $CuO_2$ planes can now be thoroughly tested in a structurally simple compound, e.g., through an extension of the present measurements to different doping levels. Our work furthermore demonstrates that the comparative quantitative study of Hg1201 and Y123 allows the determination of essential universal properties, and thus serves as a basis for a comprehensive understanding of the cuprates.

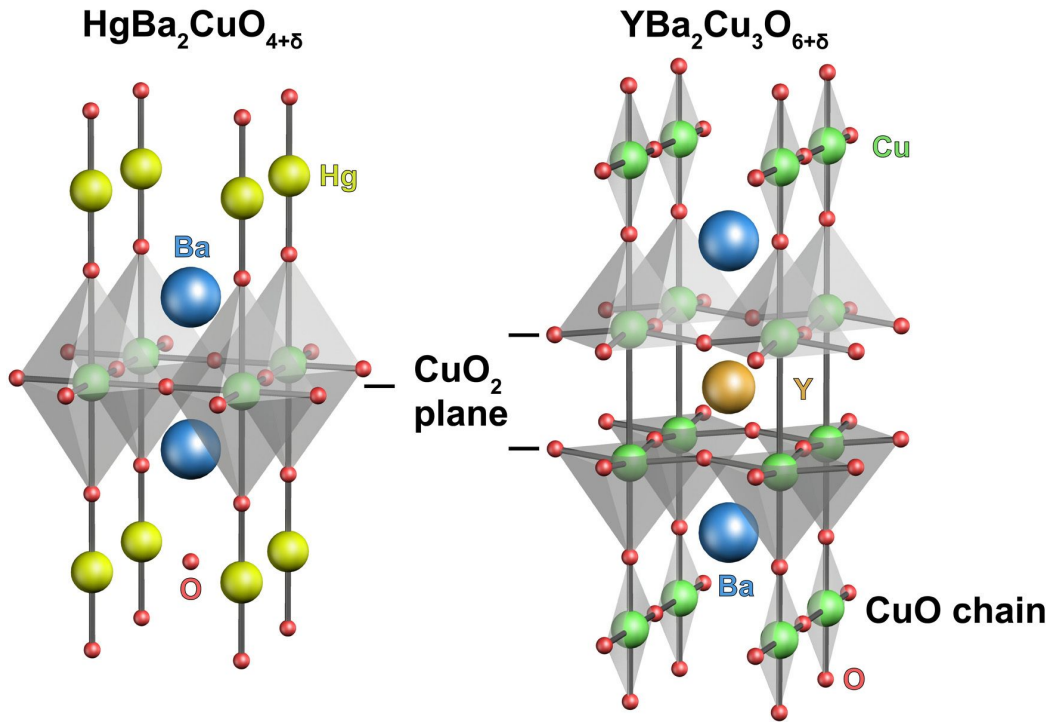

**Figure 1 | Crystal structures of Hg1201 and Y123.** The two compounds considerably differ in their structural symmetry (tetragonal versus orthorhombic), absence versus presence of Cu-O chains, number of CuO$_2$ planes per primitive cell (one versus two), CuO$_6$ octahedra versus CuO$_5$ pyramids, $c$-axis dimensions ($c \approx 9.53$ Å versus 11.65 Å), the most prevalent disorder types, etc.[30] In Hg1201, the hole concentration in the CuO$_2$ planes is altered by varying the density of interstitial oxygen atoms in the Hg layer (as indicated schematically), whereas in Y123 it is controlled by the density of oxygen atoms in the CuO chains (in underdoped Y123, a significant fraction of chain O sites is vacant; not shown).

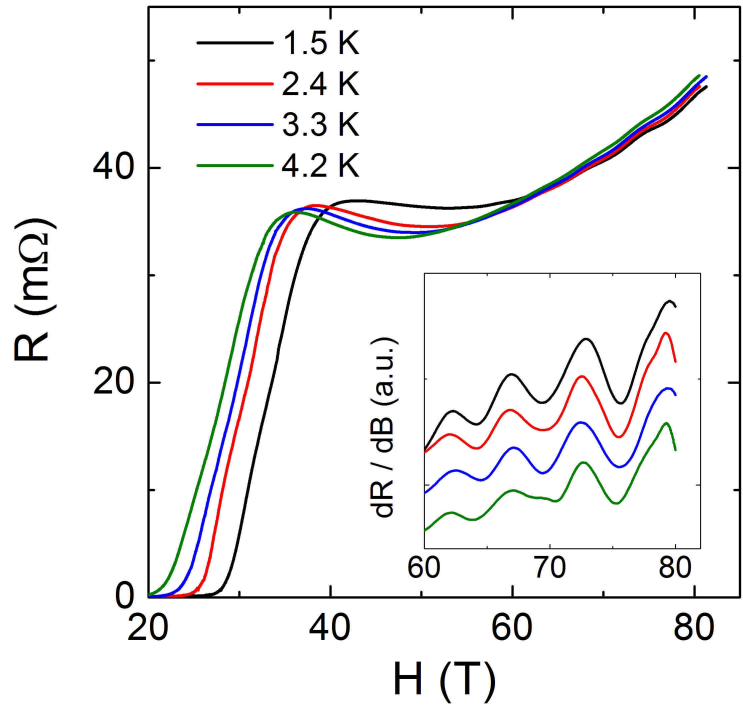

**Figure 2 | Quantum oscillations in Hg1201.** Planar transverse magnetoresistivity data for a Hg1201 sample (zero-field $T_c$ = 72 K, $p \approx 0.09$) in pulsed magnetic fields (up to 80 T) at selected temperatures (1.5 K, 2.4 K, 3.3 K and 4.2 K). QO are clearly observed from the raw data above 60 T. Inset: Derivative of the resistivity highlights the oscillations.

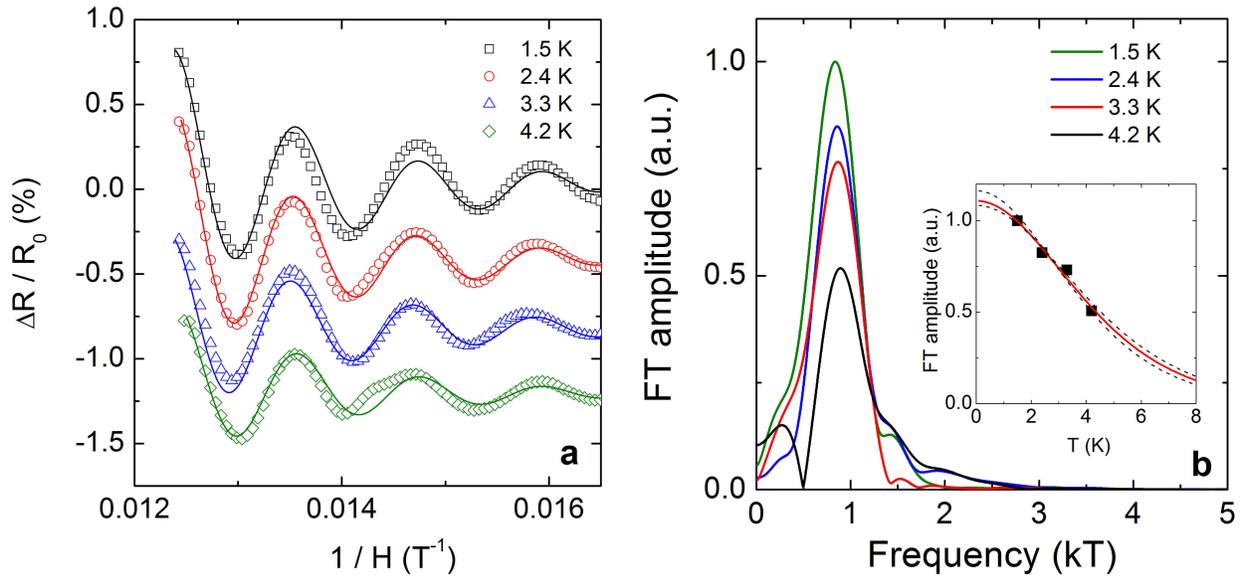

**Figure 3 | Frequency of oscillations and effective mass.** a, Oscillatory part of the isothermal magnetoresistivity data (obtained by subtracting a monotonic contribution) versus inverse field (symbols). Solid lines correspond to Lifshitz-Kosevich fit (see text). b, Fourier transform of the oscillatory part, from which only one peak is observed at F = 840 ± 30 T, with temperature-independent position. Inset: In accordance with the Lifshitz-Kosevich formula, the temperature dependence of the amplitude yields a cyclotron mass of $m^* = (2.45 \pm 0.15)\,m_e$, where $m_e$ is the mass of a free electron. Solid red line: fit. Black dashed lines: error range. See SI for details of the analysis.

|  | $T_c$ (K) | $p$ | $F$ (T) | $p_{eff}$ | $m^*$ ($m_e$) | *Hall sign* |
|---|---|---|---|---|---|---|
| Hg1201 | 72 | 0.09 | 840 ± 30 | 0.061 ± 0.002 | 2.45 ± 0.15 | (−)[28] |
| Y123 [12] | 57.5 | 0.10 | 530 ± 20 | 0.038 ± 0.002 | 1.9 ± 0.1 | (−)[28] |
| Y124 [13] | 81 | 0.14 | 660 ± 30 | 0.047 ± 0.002 | 2.7 ± 0.3 | (−)[13] |
| Tl2201[21] | 15 | 0.30 | 18 100 ± 50 | $1 + p = 1.3$ | 5 ± 1 | (+)[15] |

**Table 1 | Evolution of the Fermi surface.** Quantum oscillations have been observed in underdoped Hg1201, Y123 and Y124, as well as in overdoped Tl2201. The first two columns list the respective superconducting transition temperatures ($T_c$) and doped concentrations $p$ (per planar copper atom). Coefficients extracted from quantum oscillation experiments are: $F$ – oscillation frequency; $p_{eff}$ - carrier density per pocket; $m^*$ - cyclotron mass, where $m_e$ is the free electron mass. The last column gives the sign of the Hall coefficient measured in high fields and low temperature. In underdoped compounds, the carrier density per pocket, $p_{eff}$, is substantially smaller than that expected from band structure calculations $(1 + p)$.

**Supplementary Information** accompanies the paper on www.nature.com/naturephysics.

Correspondence should be addressed to N.B. (nbarisic@ifs.hr), C.P. (cyril.proust@lncmi.cnrs.fr) and M.G. (greven@physics.umn.edu).

**Acknowledgements**

We thank S. Barišić, L. P. Gor'kov, M.-H. Julien, D. van der Marel, S. Sachdev, L. Taillefer for useful discussions. N.B. acknowledges F. Rullier-Albenque for sharing her insights into the transport of the cuprates. The work at the University of Minnesota was supported by the Department of Energy, Office of Basic Energy Sciences. The work in Toulouse was supported by the French ANR SUPERFIELD, Euromagnet II, and the LABEX NEXT. The work in Zagreb was supported by the Unity through Knowledge Fund. N.B. acknowledges support though a Marie Curie Fellowship.

**Competing Interests** The authors declare that they have no competing financial interests.

**Author Contributions** N.B., C.P. and M.G. conceived the experiment. N.B., S.B., M.K.C., B.V., J.B. and C.P. performed the high-field transport measurements. N.B., M.K.C., C.D., W.T., G.Y. and X.Z. prepared the samples (crystal growth, annealing, magnetization and transport characterization, contacts). S.B., B.V. and C.P. analyzed the data. N.B., C.P. and M.G. wrote the manuscript with input from all authors.

Supplementary Information for:

# Universal quantum oscillations in the underdoped cuprate superconductors

N. Barišić, S. Badoux, M. K. Chan, C. Dorow, W. Tabis, B. Vignolle, G. Yu, J. Béard, X. Zhao, C. Proust, M. Greven

## A - Sample preparation and experimental setup

Hg1201 single crystals were grown by a two-step flux method [1] and subsequently annealed for several weeks at 400 $^{o}$C at air pressure of 100 Torr [2], and then quenched to the room temperature. The width of the superconducting transition, as determined from magnetic susceptibility, was typically 2-3 K with a mid point at $T_c$ = 71 K. According to the $T_c$ vs. $p$ relationship [3], this corresponds to a hole concentration of $p \approx 0.09$. The samples were then cleaved to obtain fresh and clean *ac*-surfaces onto which gold pads were evaporated. Wires (20 μm diameter) were attached with silver paste, subsequently cured for several minutes under the same conditions as the prior long-term anneal, and then quenched to room temperature. The samples were then mounted on a home-made sample-holder and introduced into a standard $^4$He cryostat that enables measurements at fixed temperatures between 1.5 K and 300 K. Resistivity measurements were performed in a conventional four-points configuration, with a current excitation of 20 mA at a frequency of ~ 60 kHz. A high speed acquisition system was used to digitize the reference signal (current) and the voltage drop across the sample at a frequency of 500 kHz. The data were analyzed with software that performs the phase comparison. Data for the rise and fall of the field pulse were in good agreement, thus we can safely exclude any heating due to Eddy currents.

## B - Coil description

In order to obtain a high signal-to-noise ratio in pulsed magnetic fields, long pulse durations are essential to significantly reduce vibrations, Eddy currents and related sample heating. The dual-coil magnet developed at the LNCMI-Toulouse produces non-destructive magnetic fields up to 80 T and provides pulse durations of more than 10 ms above 70 T (Fig. S1). Both coils are based on the optimized density reinforcement technique [4]. The outer magnet is energized with the 14 MJ capacitor bank (the main pulsed energy source of the LNCMI-Toulouse) and produces 33 T in a 170 mm bore. The conductor used is 26 mm² Hoganas

Glidcop Al-15 insulated with Kapton and reinforced with Toyobo Zylon fibres. In the middle of the winding, a cooling channel is designed where liquid nitrogen can flow. This coil weights more than 250 kg. The inner magnet produces 50 T in a 13 mm bore and is energized with a mobile 1.15 MJ capacitor bank. The conductor consists of a 6 mm² copper/stainless steel macro-composite, fabricated in the laboratory, and is also insulated with Kapton and reinforced with Zylon. With this dual coil magnet, the wait time between 80 T pulses is 90 minutes.

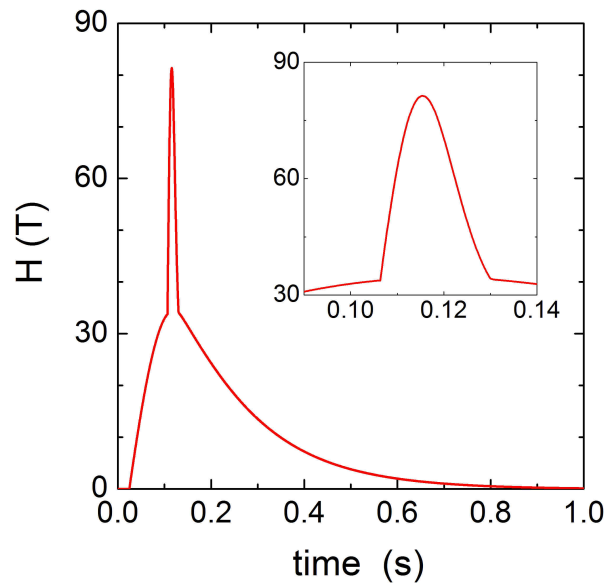

**Figure S1 | Magnetic field profile.** Time dependence of the magnetic field during one pulse as measured by an in-situ pick-up coil. Inset: field profile around the maximum field.

## C - Reproducibility

Quantum oscillations were observed in a second sample with similar doping level, but with somewhat lesser resolution. As shown in Fig. S2, the magnetoresistivity of this sample is about four times smaller than that of the sample discussed in the main text. The inset of Fig. S2 shows that the field derivatives of the magnetoresistance exhibit very similar quantum oscillation frequencies for both samples.

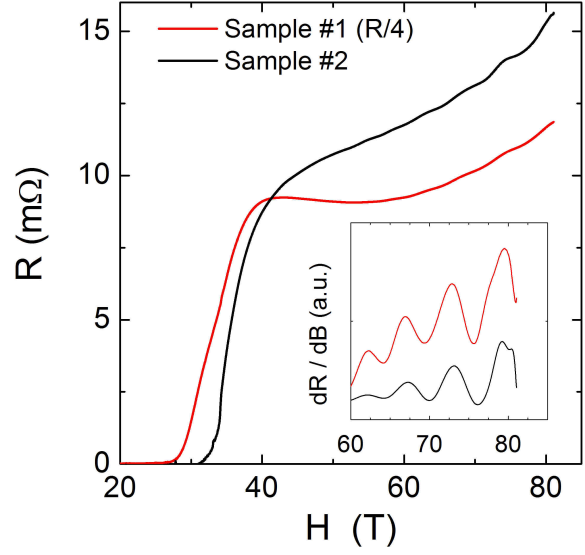

**Figure S2 | Quantum oscillations in a second sample.** Comparison of the field dependence of the resistance for the two samples that were investigated. The resistive signal of sample #1 discussed in the main manuscript (red line) has been divided by a factor of four. The inset displays the field derivative of the resistance for both samples.

## D - Determination of physical properties

The Onsager relation $F = \frac{\Phi_0}{2\pi^2} A_k$, where $\Phi_0$ is the magnetic flux quantum, allows the estimation of the cross-sectional area of the Fermi surface: the measured frequency $F = 840$ T corresponds to $A_k = 8.0$ nm$^{-2}$. Given that the area of the first Brillouin zone is $A_{BZ} = \frac{4\pi^2}{a^2}$, where $a = 3.88$ Å is the planar lattice constant, the Fermi surface area deduced from quantum oscillations represent only about 3.0 % of the first Brillouin zone. The two-dimensional carrier density per pocket is given by $n_{2D} = \frac{2A_k}{(2\pi)^2} = \frac{F}{\Phi_0} = 4.05 \ 10^{17}$ m$^{-2}$. The carrier density per pocket per Cu atom is thus $p_{eff} = \frac{n_{2D}}{1/a^2} = 0.061$.

The fits shown in Fig. 3a were obtained using the Lifshitz-Kosevich (LK) formula [5]:

$$\frac{\Delta \sigma}{\sigma} \approx \frac{\Delta \rho}{\rho} \propto A_0 \sqrt{H} R_T R_D \cos\left[2\pi\left(\frac{F}{H} - \gamma\right)\right]$$

where $R_T = \frac{\alpha T m^*/H}{\sinh(\alpha T m^*/H)}$, $R_D = \exp\left(-\frac{\alpha T_D m^*}{H}\right)$, and $\alpha = 2\pi^2 m^* k_B/e\hbar$

The cyclotron mass $m^*$ was obtained from fits of the amplitude of the Fourier transform of the data with respect to $R_T$. The Dingle temperature $T_D$ was obtained by fitting the oscillatory part of the magnetoresistance to the LK formula. A fit of the LK equation to the data in Fig. 3a gives a single frequency $F$ of $840 \pm 30$ T but given the small amplitude of the oscillation, we cannot rule out the presence of other frequency.